# Quantum Mechanics as Electrodynamics of Curvilinear Waves

**Alexander G. Kyriakos**

*Saint-Petersburg State Institute of Technology,
St.Petersburg, Russia*

Present address:
   *Athens, Greece
E-mail: agkyriak@otenet.gr*

**Abstract**

The suggested theory is the new quantum mechanics (QM) interpretation. The below research proves that the QM represents the electrodynamics of the curvilinear closed (non-linear) waves. It is entirely according to the modern interpretation and explains the particularities and the results of the quantum field theory.

**Contents**

**1. INTRODUCTION**

The attempts to explain the quantum theory and the particle structure were already undertaken in the last century (Lord Kelvin, H.A.Lorentz, Lois de Broglie et.al.) and are continued until today [1].

In the following article we show that the quantum mechanics is mathematically isomorphic to the electrodynamics of the curvilinear wave. The basis and the results of the theory consist in the following:
1) It is accepted as a consequence of Maxwell's theory and the uncertainty relation of Heisenberg that the photon represents a classical relativistic string, which has one wave length.
2) It is accepted as a hypothesis that the reason for the formation of the particles is the movement of a photon along the closed curvilinear trajectory.
3) It is shown that the electron and the positron appears as the consequence of the division of the twirled photon into two completely antisymmetric twirled semi-photons.
4) It is shown that to this corresponds the decomposition of the wave equation of a photon into two Dirac's equations about the electron and the positron.
5) It is shown that the reason of the occurrence of the currents (charges) particles is the movement of the electric vector of a photon on a curvilinear trajectory (that corresponds to the occurrence of the additional members in the linear equations).
6) It is shown that the equations of the particle structure are non-linear and this nonlinearity is of the same type, as in the equations of the theory of Standard Model.
7) It is shown that the twirled string (photon) has an integer spin (i.e. it is a boson), and the twirled semi-photon has a half-integer spin (i.e. it is a fermion).
8) It is shown that all parities of QM correspond precisely to the electrodynamics of the curvilinear waves and have formally the same kind, as in Maxwell's theory.
9) It is also shown that all the postulates of QM (for example, the statistical interpretation of the wave function, dualism a wave – particle, a principle of conformity, sense of the Dirac's matrix representation, and so forth), and also all the mathematical particularities of Dirac's theory follow from the electromagnetic form of the theory, as consequences.

Thus, the below research proves *that the QM represents the electrodynamics of the curvilinear closed (non-linear) waves*.

The article contains two parts: A. the quasi-classical theory, and B. the exact theory.

*A. QUASI-CLASSICAL THEORY.* In the world of the elementary particles the following approximate electron model theory is the same as the hydrogen planetary model theory in the atom theory.

**2. BASIC HYPOTHESES**

These hypotheses don't conflict with the modern theory.
**Photon hypotheses**
   1. *A photon is described by Maxwell's equations.*
   2. *All photons contain one wave period.*

**Particles production hypotheses**
   Let's consider the particle-antiparticle production conditions.

One $\gamma$-quantum cannot turn spontaneously into the electron-positron pair, although it interacts with the electron-positron vacuum. For the pair production, at first, the following mass correlation is necessary: $\varepsilon_p \geq 2m_e c^2$ (where $\varepsilon_p$ is the photon energy, $m_e$ is the electron mass and $c$ is the light velocity). At second, the presence of the other particle, having the electromagnetic field, is needed. For example, we have the typical reaction:

$$\gamma + N \to N + e^+ + e^-,$$

which means that, while moving through the nucleus field $N$ the photon takes some transformation, which corresponds to the pair production. Considering the fact that Pauli's matrices describe the vector space rotations and taking in account the optical-mechanical analogy analysis also, we can assume that the above transformation is a field distortion. From this follows the **distortion hypothesis:**

*By the fulfilment of the pair production conditions, the distortion of the electromagnetic field of photon can take place; as result photon is able to move along the closed trajectory, making some stable construction named elementary particle.*

(The necessity to introduce some other hypothesis, **division hypothesis**, will be understood from the content of the following chapters).

### 3.0 ELECTROMAGNETIC MODEL OF ELECTRON

According to the modern experiments and ideas, the leptons don't have a structure. That's the reason why it is said that the leptons are point particles. But as it's known, it means only that the electrodynamics is right on any short distance, by modern experiments until $2 \times 10^{-18}$ m. So, this fact not only does contradict to our supposition about electron, «created» from the electromagnetic field, but also verifies it.

### 3.1. STRUCTURE AND PARAMETERS OF THE ELECTRON MODEL

According to the hypotheses the electron must be presented as the torus. It's obvious that the torus radius is equal to $r_t = \dfrac{\lambda_p}{2\pi}$, where $\lambda_p$ is the photon's wavelength.

In our case the photon characteristics are defined by the electron-positron pair production conditions: by the photon energy $\varepsilon_p = 2m_e c^2$ and by the circular frequency: $\omega_p = \dfrac{\varepsilon_p}{\hbar} = \dfrac{2m_e c^2}{\hbar}$. Therefore, the photon wavelength is $\lambda_p = \dfrac{2\pi c}{\omega_p} = \dfrac{\pi \hbar}{m_e c}$. Taking in account this value we can write: $r_t = \dfrac{\hbar}{2m_e c} = \dfrac{1}{\alpha} \dfrac{e^2}{2m_e c^2}$, where $\alpha$ is the electromagnetic constant.

Let us suppose also that the cross-section torus radius $r_c$ is equal to $r_c = \zeta\, r_t$, where $\zeta \leq 1$ (fig.1).

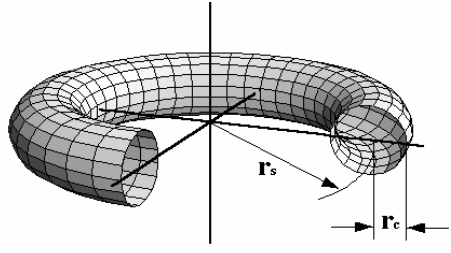

Fig.1. Electron model

## 3.2. RING CURRENT

Let the plane polarised photon, which have the fields $(E_x, H_z)$, moves along the $y$-axis and rolls up in the plane $X,O',Y$ with radius $r_t$ in the fixed co-ordinate system $X',Y',Z',O'$. It could be said, that along the tangent to the circumference the rectangular axes system $(x,-y,z)$ or $(E_x, S_y, H_z)$, where $S_y = [\vec{E} \times \vec{H}]_y$ is the Pointing vector, moves, having the $-y$ direction.

Let us show that due to the electromagnetic wave distortion by the photon twirling the displacement ring current arises.

According to Maxwell [2] the displacement current is defined by:

$$\vec{j}_{dis} = \frac{1}{4\pi} \frac{\partial \vec{E}}{\partial t}, \quad (3.1)$$

Let us calculate this current for the wave, moving along the ring. The electrical field vector $\vec{E}$ (let it have direction from the centre), which moves along the distortion trajectory, can be written in form:

$$\vec{E} = -E \cdot \vec{n}, \quad (3.2)$$

where $E = |\vec{E}|$ is the absolute value and $\vec{n}$ is the normal unit-vectors. Derivative of $\vec{E}$ with respect to $t$ can be represented as:

$$\frac{\partial \vec{E}}{\partial t} = -\frac{\partial E}{\partial t} \vec{n} - E \frac{\partial \vec{n}}{\partial t}, \quad (3.3)$$

Here the first term has the same direction, as $\vec{E}$. The existence of the second term shows that at the wave distortion the supplementary displacement current appears. It is not difficult to show that it has a direction, tangent to the ring:

$$\frac{\partial \vec{n}}{\partial t} = -\frac{v_p}{r_p} \vec{\tau}, \quad (3.4)$$

where $\vec{\tau}$ is the tangential unit-vectors, $v_p = c$ is the photon velocity, $r_p \equiv r_t$ is the radius of the twirled photon (i.e torus). Then the displacement current of the ring wave can be written:

$$\vec{j}_{dis} = -\frac{1}{4\pi} \frac{\partial E}{\partial t} \vec{n} + \frac{1}{4\pi} \omega_p E \cdot \vec{\tau}, \quad (3.5)$$

where $\omega_p = \dfrac{c}{r_p}$ - the angular velocity or angular frequency, $\vec{j}_n = \dfrac{1}{4\pi}\dfrac{\partial E}{\partial t}\vec{n}$ and $\vec{j}_\tau = \dfrac{\omega_p}{4\pi} E \cdot \vec{\tau}$ are the normal and tangent components of the twirled photon current correspondingly. So:

$$\vec{j}_{dis} = \vec{j}_n + \vec{j}_\tau, \quad (3.6)$$

The currents $\vec{j}_n$ and $\vec{j}_\tau$ are always mutually perpendicular, so that we can write in complex form:

$$j_{dis} = j_n + i j_\tau,$$

where $i = \sqrt{-1}$. As we see, *the appearance of an imaginary unity in the quantum mechanics is tied with the tangent current.*

The same results can be obtained by using the general methods of the distortion field investigation (see below chapter 4.2.).

## 3.3. CHARGE APPEARANCE AND DIVISION HYPOTHESIS

It's not difficult to calculate now the charge density of the twirled photon:

$$\rho_p = \dfrac{j_\tau}{c} = \dfrac{1}{4\pi}\dfrac{\omega_p}{c} E = \dfrac{1}{4\pi}\dfrac{1}{r_p} E, \quad (3.7)$$

The full charge of the particle (i.e. of the twirled photon) can be defined by integrating along all the torus volume $\Delta\tau_p \equiv \Delta\tau_t$:

$$q = \int_{\Delta\tau_t} \rho_p d\tau, \quad (3.8)$$

Using the model (fig.1) and taking $\vec{E} = \vec{E}(l)$, where $l$ is the length of the way, we obtain:

$$q = \int_{S_t}\int_0^{\lambda_p} \dfrac{1}{4\pi}\dfrac{\omega_p}{c} E_o \cos k_p l \, dl \, ds = \dfrac{1}{4\pi}\dfrac{\omega_p}{c} E_o S_c \int_0^{\lambda_p} \cos k_p l \, dl = 0, \quad (3.9)$$

(here $E_o$ is the amplitude of the twirled photon wave field, $S_c$ - the area of torus cross-section, $ds$ is the element of the surface, $dl$ - the element of the circle length, $k_p = \dfrac{\omega_p}{c}$ is the the value of the wave-vector).

It's easy to understand these results: because the ring current changes its direction every half-period, the full charge is equal to zero.

Therefore, *the above model may represent only the non-charged particle*. It's clear that a charge particle, must contain only one half-period of wave.

For the solution of this problem we suggest the following **division hypothesis:** *at the moment when the photon begins to roll up, the spontaneous photon division in two half-periods is possible* (see fig.2).

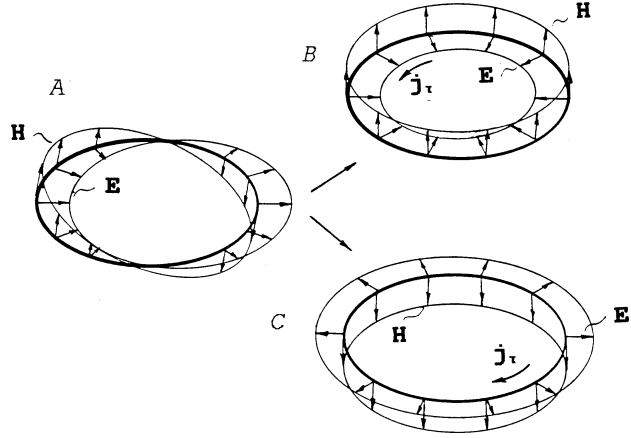

Fig.2 Twirled photon division

It is clear that *the parts B and C* (fig.2) *contain the currents of opposite directions.*

We can suppose, that *the cause of a photon's division is the mutual repulsion of oppositely directed currents.*

It's also clear that *the division process corresponds to the process of the particle-antiparticle pair production.* It's proved out by the fact, that the daughter's twirled semi-photons B and C are completely anti-symmetric and can't be transformed one to each other by any transformation of co-ordinates (if it is not to be accompanied by the changing of field direction).

It is interesting that both the twirled semi-photon and the twirled photon radii must be the same. This fact follows from the momentum conservation low. Really, the twirled photon momentum is equal to:

$$\sigma_p = p_p \cdot r_p = 2m_e c \cdot \frac{\hbar}{2m_e c} = 1\hbar ,  \qquad (3.10)$$

In accordance with the momentum conservation low:

$$\sigma_s^+ + \sigma_s^- = \sigma_p , \qquad (3.11)$$

where $\sigma_s^+, \sigma_s^-$ are the spins of the plus and minus semi-photons (i.e. of the electron and positron). Then we obtain:

$$\sigma_s = \frac{1}{2}\sigma_h = \frac{1}{2}\hbar , \qquad (3.12)$$

Since

$$\sigma_s = p_s \cdot r_s , \qquad (3.13)$$

where $r_s$ is the twirled semi-photon (electron) radius, and $p_s = m_e c$ is the inner semi-photon (electron) impulse, we have

$$r_s = \frac{\sigma_s}{p_s} = \frac{1}{2}\frac{\hbar}{m_e c} = \frac{\hbar}{2 m_e c} = r_p \equiv r_t, \quad (3.14)$$

So, *the torus size doesn't change after division.*
The angular rotation velocity (angular frequency) also doesn't change and $\omega_s = \frac{c}{r_s} = \frac{2 m_e c^2}{\hbar} = \omega_p$. From this it also follows that the volumes and areas of the twirled photon and the semi-photon will be the same: $\Delta\tau_s = \Delta\tau_p$, $S_s = S_c$

The division hypothesis makes it possible to outline the solution of the some fundamental problems:

    1) *the origin of the charge conservation low*: since in nature there are the same numbers of the photon semi-periods of positive and negative directions the sum of the particles charge is equal zero.

    2) *the difference between positive and negative charges*: this difference follows from the difference of the field and tangent current direction of the twirled semi-photons after pair production (by condition that the Pauly -principle is true).

    3) *Zitterbewegung*. The results obtained by E.Schroedinger in his well-known articles about the relativistic electron [3] are the most important confirmation for the electron structure model. He showed, that electron has a special inner motion "Zitterbewegung", which has frequency $\omega_Z = \frac{2 m_e c^2}{\hbar}$, amplitude $r_Z = \frac{\hbar}{2 m_e c}$, and velocity of light $\upsilon = c$. The attempts to explain this motion had not given results (see e.g. [4,5]). But if the electron is the twirled semi-photon, then we receive the simple explanation of Schroedinger's analysis.

    4) *the difference between bosons and fermions* is well explained: the bosons contain the even number and the fermions contain the odd number of the twirled semi-photons.

    5) *Infinite problems*: in the above theory the charge and mass infinite problems don't exist.

### 3.4. CHARGE AND FIELD MASS OF TWIRLED SEMI-PHOTON

We can now calculate the semi-photon charge:

$$q = \frac{1}{\pi}\frac{\omega_s}{c} E_o S_s \, 2\int_0^{\frac{\lambda_s}{4}} \cos k_s l \, dl = \frac{1}{\pi} E_o S_c, \quad (3.15)$$

Since $S_s = \pi r_c^2$, we obtain:

$$q = E_o r_c^2 = \zeta^2 E_o r_s^2, \quad (3.16)$$

To calculate the mass we must calculate first the energy density of the electromagnetic field:

$$\rho_\varepsilon = \frac{1}{8\pi}\left(\vec{E}^2 + \vec{H}^2\right), \quad (3.17)$$

and we have $|\vec{E}| = |\vec{H}|$ in Gauss's system. Therefore, Eq.(3.17) can be written so:

$$\rho_\varepsilon = \frac{1}{4\pi} E^2, \quad (3.18)$$

Using (3.18) and a well-known relativistic relationship between a mass and energy densities:

$$\rho_m = \frac{1}{c^2}\rho_\varepsilon , \qquad (3.19)$$

we obtain:

$$\rho_m = \frac{1}{4\pi c^2}E^2 = \frac{1}{4\pi c^2}E_o^2 \cos^2 k_s l , \qquad (3.20)$$

Using (3.20), we can write for the semi-photon mass (fig.1):

$$m_s = \iint_{S_t, l}\rho_m ds\, dl = \frac{S_c E_o^2}{\pi c^2}\int_0^{\frac{\lambda_s}{4}}\cos^2 k_s l\, kl , \qquad (3.21)$$

From (3.21), calculating the integral, we obtain the semi-photon mass:

$$m_s = \frac{E_o S_c}{4\omega_s c} = \frac{\pi \zeta^2 E_o^2 r_s^2}{4\omega_s c} , \qquad (3.22)$$

### 3.5. ELECTROMAGNETIC CONSTANT

Using Eqs. (3.16) and (3.22) we can write:

$$m_s = \frac{\pi q^2}{4\zeta^2 \omega_s c r_s^2} , \qquad (3.23)$$

or, taking in account that $\omega_s \cdot r_s = c$ we obtain:

$$r_s = \frac{\pi}{2\zeta^2}\frac{q^2}{2m_s c^2} , \qquad (3.24)$$

Putting here the values $\omega_s$ and $r_s$, we have:

$$\frac{q^2}{\hbar c} = \frac{2}{\pi}\zeta^2 = \alpha_q \approx 0{,}637\zeta^2 , \qquad (3.25)$$

which corresponds to the electromagnetic constant $\frac{e^2}{\hbar c} = \alpha \cong \frac{1}{137}$.

It is known that the ratio of the electron classical radius value

$$r_o = \frac{e^2}{2mc^2} , \qquad (3.26)$$

to the electron quantum radius value (i.e. to the Compton electron wave length) $r_C = \frac{\hbar}{2mc} \equiv r_s$ is equal to the fine structure constant $\frac{r_o}{r_s} = \alpha$. As long as in our calculations we didn't take into account the physical vacuum polarization, it would be right to consider the $r_s$ as the "bare" electron radious. Then we can suppose that $r_o$ is the real electron radius.

Although the constant $\alpha_q$ has the indeterminate value $\zeta$, its calculation should be understood as a theory success. The formula (3.25) shows that in our theory the electric charge is defined only by the universe constants; that means that there are no free charges less than this one. At the same time, our theory doesn't limit the mass value of twirled semi-photon; this fact is also according to the experimental data.

## 3.6. STABILITY OF THE ELECTRON MODEL

This problem is very important for the existence of our model. In the classical electron theory the forces, which can keep the parts of charge together, had not been found. In our theory such forces exist. The magnetic component of Lorentz's force, which acts in opposite direction to the Coulomb's electric force, appears in the twirled semi-photon.

In fact, as result of the current $\vec{j}_\tau$ and magnetic field $\vec{H}$ interaction, the magnetic force density appears:

$$\vec{f}_M = \frac{1}{c}\left[\vec{j}_\tau \times \vec{H}_s\right], \quad (3.27)$$

It is difficult to say something more about the electron model field distribution here (note only that the twirled semi-photon fields are like the tokamak fields). In this area there is a number of achievements both old authors and new (see e.g.[6,7]).

Here we shall be satisfied by the fact that the electron stability is the consequence of the exact theory (see below).

## 3.7. SPIN OF MODEL

Using the data of the torus model we can calculate the spin of the twirled semi-photon:
$$\sigma_s = p_s \cdot r_s = \frac{1}{2}\hbar, \quad (3.28)$$

## 3.8. MAGNETIC MOMENT OF MODEL

Magnetic moment accordingly with definition is:
$$\mu_s = I \cdot S_I, \quad (3.29)$$

where $I$ is electron ring current and $S_I$ is the current ring square.
In our case we have:

$$I = q\frac{\omega_s}{2\pi} = q\frac{1}{2\pi}\frac{2m_e c^2}{\hbar}, \quad (3.30)$$

$$S_I = \pi\, r_s^{\,2} = \pi\left(\frac{\hbar}{2m_e c}\right)^2, \quad (3.31)$$

Using these formulae, we find:

$$\mu_s = \frac{1}{2}\frac{q\hbar}{2m_e}, \quad (3.32)$$

If we put $q = e$, the value (3.32) is equal to half of the experimental value of the magnetic momentum of the electron. Taking into account Thomas's precision we obtain the experimental value of the electron magnetic momentum.

*B. EXACT THEORY*. It is entirely according to the modern theory.

## 4. LINEAR ELECTRODYNAMICS' AND QUANTUM EQUATIONS OF TWIRLED SEMI-PHOTON

### 4.1. ELECTRODYNEMICS' FORM OF DIRAC'S EQUATION

The twirled semi-photons, as fermions, must be defined by the spinors, which have other transformation properties than the vectors of Maxwell's electromagnetic field. On the other hand the theory of elementary particles, describing the non-linear wave, must be non-linear.

Therefore, *the Dirac equations must be the limit of the non-linear equations and some modification of Maxwell's equations*. Let us prove it.

Consider the linear photon, moving along $y$-axis. In a more general case it has the two possible polarisations and contains the field vectors $(E_x, E_z, H_x, H_z)$ $(E_y = H_y = 0)$. Such photon can form the ring only on the $X,O,Y$ or $Y,O,Z$ plain.

The bispinor form of Dirac's equations can be written as one equation [8]:

$$\hat{\varepsilon}\psi + c\hat{\vec{\alpha}}\,\hat{\vec{p}}\psi + \hat{\beta}\,mc^2\psi = 0, \qquad (4.1)$$

where $\hat{\vec{\alpha}}, \hat{\beta}$ - are Dirac's matrices by [8]:

$$\hat{\alpha}_0 = \begin{pmatrix} 1 & 0 & 0 & 0 \\ 0 & 1 & 0 & 0 \\ 0 & 0 & 1 & 0 \\ 0 & 0 & 0 & 1 \end{pmatrix}, \quad \hat{\alpha}_1 = \begin{pmatrix} 0 & 0 & 0 & 1 \\ 0 & 0 & 1 & 0 \\ 0 & 1 & 0 & 0 \\ 1 & 0 & 0 & 0 \end{pmatrix},$$

$$\hat{\alpha}_2 = \begin{pmatrix} 0 & 0 & 0 & -i \\ 0 & 0 & i & 0 \\ 0 & -i & 0 & 0 \\ i & 0 & 0 & 0 \end{pmatrix}, \quad \hat{\alpha}_3 = \begin{pmatrix} 0 & 0 & 1 & 0 \\ 0 & 0 & 0 & -1 \\ 1 & 0 & 0 & 0 \\ 0 & -1 & 0 & 0 \end{pmatrix}, \quad \vec{\alpha}_4 \equiv \hat{\beta} = \begin{pmatrix} 1 & 0 & 0 & 0 \\ 0 & 1 & 0 & 0 \\ 0 & 0 & -1 & 0 \\ 0 & 0 & 0 & -1 \end{pmatrix}$$

$\hat{\varepsilon} = i\hbar\dfrac{\partial}{\partial t}$, $\hat{\vec{p}} = -i\hbar\vec{\nabla}$ are the operators of energy and momentum, $\psi$ is the wave function named bispinor, which has four component.

Put the following semi-photon bispinor:

$$\psi = \begin{pmatrix} \psi_1 \\ \psi_2 \\ \psi_3 \\ \psi_4 \end{pmatrix} = \begin{pmatrix} E_x \\ E_z \\ iH_x \\ iH_z \end{pmatrix}, \quad \psi^+ = \begin{pmatrix} E_x & E_z & -iH_x & -iH_z \end{pmatrix}, \qquad (4.2)$$

Taking into account that $\psi = \psi(y)$, from (4.1) we obtain:

$$\begin{cases} rot\,\vec{E} + \dfrac{1}{c}\dfrac{\partial \vec{H}}{\partial t} = i\dfrac{\omega}{c}\vec{H}, \\ \\ rot\,\vec{H} - \dfrac{1}{c}\dfrac{\partial \vec{E}}{\partial t} = i\dfrac{\omega}{c}\vec{E}, \end{cases} \qquad (4.3)$$

Eqs. (4.3) are the Maxwell equations with the imaginary currents. It is interesting that together with the electrical current the magnetic current also exists here.

This current is equal to zero by Maxwell's theory, but its existence by Dirac doesn't contradict to the quantum theory. (According to the full theory the magnetic current appearance is related to the origin photon circular polarization).

So, our first suppositions are right. Now we must show that all the quantum mechanics can have the electrodynamics' form.

Note also that the formal matrix form of the Maxwell's equations, known for a long time, acquires in our theory a physical sense.

## 4.2. ELECTRODYNAMICS' AND QUANTUM FORM OF ELECTRON-POSITRON PAIR PRODUCTION THEORY

Using Eq. (4.2), we can write the equation of the electromagnetic wave moved along the $y$-axis in form:

$$\left(\hat{\varepsilon}^2 - c^2 \hat{\vec{p}}^2\right)\psi = 0, \qquad (4.4)$$

The Eq. (4.4) can also be written in the following form:

$$\left[\left(\hat{\alpha}_o \hat{\varepsilon}\right)^2 - c^2\left(\hat{\vec{\alpha}} \, \hat{\vec{p}}\right)^2\right] \psi = 0, \qquad (4.5)$$

In fact, taking into account that

$$\left(\hat{\alpha}_o \hat{\varepsilon}\right)^2 = \hat{\varepsilon}^2, \quad \left(\hat{\vec{\alpha}} \, \hat{\vec{p}}\right)^2 = \hat{\vec{p}}^2,$$

we see that Eqs.(4.4) and (4.5) are equivalent.

Factorising Eq. (4.5) and multiplying it from left on Hermithian-conjugate function $\psi^+$ we get:

$$\psi^+ \left(\hat{\alpha}_o \hat{\varepsilon} - c\hat{\vec{\alpha}} \, \hat{\vec{p}}\right)\left(\hat{\alpha}_o \hat{\varepsilon} + c\hat{\vec{\alpha}} \, \hat{\vec{p}}\right)\psi = 0, \qquad (4.6)$$

The Eq. (4.6) may be disintegrated on two equations:

$$\psi^+ \left(\hat{\alpha}_o \hat{\varepsilon} - c\hat{\vec{\alpha}} \, \hat{\vec{p}}\right) = 0, \qquad (4.7)$$

$$\left(\hat{\alpha}_o \hat{\varepsilon} + c\hat{\vec{\alpha}} \, \hat{\vec{p}}\right)\psi = 0, \qquad (4.8)$$

It is not difficult to show (using Eq. (4.2)) that the Eqs. (4.7) and (4.8) are Maxwell's equations without current and, at the same time, are Dirac's electron-positron equations without mass.

In accordance with our assumption, the reason for current appearance is the electromagnetic wave motion along a curvilinear trajectory. We showed the current appearance in the vector form (see chapter 3.2.). The same results can be obtained, using the general methods of the distortion field investigation [9]. The question is about the tangent space introduction at every point of the curvilinear space, in which the orthogonal axis system moves. This corresponds to the fact, that the wave motion along a circular trajectory is accompanied by the motion of the rectangular basis, built on vectors ($\vec{E}, \vec{S}, \vec{H}$), where $\vec{S}$ is the Pointing vector.

For the generalisation of Dirac,s equation in Riemann's geometry [9] it is necessary to replace the usual derivative $\partial_\mu / \partial \xi_\mu$, where $\xi_\mu = \frac{1}{i\hbar c}\{-ct, x, y, z\}$, with the covariant derivative: $D_\mu^{\ \nu} = \partial_\mu + \Gamma_\nu$

($\mu, \nu = 0, 1, 2, 3$ are the summing indexes), where $\Gamma_\nu$ is the analogue of Christoffel's symbols. When a spinor moves along the line, all $\Gamma_\nu = 0$, and we have a usual derivative. If a spinor moves along the curvilinear trajectory, then not all the $\Gamma_\nu$ are equal to zero and a supplementary term appears. Typically, the last one is not the derivative, but is equal to the product of the spinor itself with some coefficient $\Gamma_\nu$. Thus we can assume that the supplementary term a longitudinal field is, i.e. it is a current.

As the increment in spinor $\Gamma_\nu$ is a 4-vector and has here the energy-momentum dimension [9] it is logical (taking into account the above results) to identify $\Gamma_\nu$ with 4-vector of energy-momentum of the own electron field:

$$\Gamma_\nu = \{\varepsilon_p, c\vec{p}_p\}, \tag{4.9}$$

Then Eqs. (4.7) and (4.8) in the curvilinear space will be have a view:

$$\psi^+[\,(\hat{\alpha}_o \hat{\varepsilon} - c\hat{\vec{\alpha}}\ \hat{\vec{p}}) - (\hat{\alpha}_o \varepsilon_s - c\hat{\vec{\alpha}}\ \vec{p}_s)\,] = 0, \tag{4.10}$$

$$[\,(\hat{\alpha}_o \hat{\varepsilon} + c\hat{\vec{\alpha}}\ \hat{\vec{p}}) + (\hat{\alpha}_o \varepsilon_s + c\hat{\vec{\alpha}}\ \vec{p}_s)\,]\psi = 0, \tag{4.11}$$

According to the energy conservation law we can write:

$$\hat{\alpha}_o \varepsilon_s \pm c\hat{\vec{\alpha}}\ \vec{p}_s = \mp\hat{\beta}\, m_e c^2, \tag{4.12}$$

Substituting Eq.(4.12) in Eqs.(4.10) and (4.11) we will arrive at the usual kind of Dirac's equation with the mass:

$$\psi^+[\,(\hat{\alpha}_o \hat{\varepsilon} - c\hat{\vec{\alpha}}\ \hat{\vec{p}}) - \hat{\beta}\, m_e c^2\,] = 0, \tag{4.13}$$

$$[\,(\hat{\alpha}_o \hat{\varepsilon} + c\hat{\vec{\alpha}}\ \hat{\vec{p}}) + \hat{\beta}\, m_e c^2\,]\psi = 0, \tag{4.14}$$

Therefore, our supposition is correct: *the electron is the twirled semi-photon.*

Note *that the alternative set of the electromagnetic field vectors corresponds to the alternative set of Dirac's matrices.*

Note also: *because $E_y = H_y = 0$ is true by any transformation, there is Dirac's electron equation property that the bispinor has only four components.*

## 4.3. ELECTRODYNAMICS' SENSE OF BILINEAR FORMS AND STATISTICAL INTERPRETATION OF THE QUANTUM MECHANICS

It is known that there are 16 Dirac's matrices of 4x4 dimensions. (We use the set of matrices by [8], which we will name $\alpha$-set). Enumerate main Dirac's matrices:

1) $\hat{\alpha}_4 \equiv \hat{\beta},$  2) $\hat{\alpha}_\mu = \{\hat{\alpha}_0, \hat{\vec{\alpha}}\} \equiv \{\hat{\alpha}_0, \hat{\alpha}_1, \hat{\alpha}_2, \hat{\alpha}_3\}$  3) $\hat{\alpha}_5 = \hat{\alpha}_1 \cdot \hat{\alpha}_2 \cdot \hat{\alpha}_3 \cdot \hat{\alpha}_4$

where 1) is a scalar, 2) is a 4-vector, 3) is a pseudoscalar. Using (4.3) and taking in account that $\psi = \psi(y)$ and $\psi^+ = (E_x\ \ E_z\ \ -iH_x\ \ -iH_z)$ (where (+) is the Hermithian conjugation sign) it is easy to obtain the electrodynamics' expression for the bilinear forms of these matrices:

1) $\psi^+ \hat{\alpha}_4 \psi = (E_x^2 + E_z^2) - (H_x^2 + H_z^2) = \vec{E}^2 - \vec{H}^2 = 8\pi\, I_1$, where $I_1$ is the first scalar of Maxwell's theory;

2) $\psi^+\hat{\alpha}_o\psi = \vec{E}^2 + \vec{H}^2 = 8\pi\ U$, and $\psi^+\hat{\alpha}_y\psi = -8\pi\ c\vec{g}_y$. As it is known, the value $\left\{\frac{1}{c}U, \vec{g}\right\}$ is the energy-momentum 4-vector.

3) $\psi^+\hat{\alpha}_5\psi = 2(E_xH_x + E_zH_z) = 2(\vec{E}\cdot\vec{H})$ is the pseudoscalar of the electromagnetic field. As it's know $(\vec{E}\cdot\vec{H})^2 = I_2$ is the second scalar of the electromagnetic field theory.

From Dirac's equation the continuity equation for the probability can be obtained:

$$\frac{\partial P_{pr}(\vec{r},t)}{\partial t} + div\ \vec{S}_{pr}(\vec{r},t) = 0, \qquad (4.15)$$

Here $P_{pr}(\vec{r},t) = \psi^+\hat{\alpha}_0\psi$ is the probability density, and $\vec{S}_{pr}(\vec{r},t) = -c\psi^+\hat{\alpha}\psi$ is probability-flux density. Using the above results we can obtain: $P_{pr}(\vec{r},t) = 8\pi\ U$ and $\vec{S}_{pr} = c^2\vec{g} = 8\pi\ \vec{S}$, where $\vec{S}$ is Pointing's vector. Taking in account the above results from (4.15-18) we get the equation:

$$\frac{\partial U}{\partial t} + div\ \vec{S} = 0, \qquad (4.16)$$

which is the energy (mass) conservation law of the electron electromagnetic field. Therefore, the Eq. (4.15) has the classical sense too.

*So, the non-normalised probability density and probability-flux density are the energy (mass) density and energy (mass)-flux density of the electromagnetic field correspondingly.* Really, normalising to unit the energy distribution we come to the distribution of probability density of the one and other value. For free immobile electron the normalisation means: $\int_0^\infty U d\tau = m_e c^2$, or $\frac{1}{8\pi\ m_e c^2}\int P d\tau = 1$, where $P$ is non-normalised probability density. Using the change: $\psi \to \sqrt{8\pi\ m_e c^2}\ \psi'$, where $\psi'$ is the normalised wave function, from the last equation we have correlation: $\int \psi'^+\psi' = 1$, which coincides with the quantum normalisation.

It is not difficult to understand the method of the quantum values calculations as the operators' eigenvalues: *if the elementary particles are the twirled electromagnetic waves, their interaction leads to the waves interference. In such case the calculation of probability density is the calculation of the energy maximums of this interference. As it is well known, in classical physics such calculation is the eigenvalue problem.*

## 4.4. LORENTZ'S FORCE

The expression of Lorentz's force by the energy-momentum tensor of electromagnetic field $\tau_\mu^\nu$ is well known [2]:

$$f_{\mu\nu} = -\frac{1}{4\pi}\frac{\partial \tau_\mu^\nu}{\partial x^\nu} \equiv -\frac{1}{4\pi}\partial_\nu\tau_\mu^\nu, \qquad (4.17)$$

This tensor is symmetrical and has the following components:

$$\tau_{pq} = -\left(E_p E_q + H_p H_q\right) + \frac{1}{2}\delta_{pq}\left(\vec{E}^2 + \vec{H}^2\right),$$

$$\tau_{p0} = 4\pi\, S_p = \left[\vec{E}\times\vec{H}\right]_p = \frac{4\pi}{c}\left(\vec{S}\right)_p, \quad (4.18)$$

$$\tau_{00} = 4\pi\, U = \frac{1}{2}\left(\vec{E}^2 + \vec{H}^2\right),$$

where $p, q = 1,2,3$, and $\delta_{pq} = 0$ when $p = q$, and $\delta_{pq} = 1$ when $p \neq q$.
Using (4.18) it can be written:

$$f_1 = f_3 = 0, \quad f_2 \equiv -\left(\frac{\partial \vec{g}}{\partial t} + \mathrm{grad}\, U\right), \quad (4.19)$$

$$f_0 = -\left(\frac{1}{c}\frac{\partial U}{\partial t} + c\, \mathrm{div}\, \vec{g}\right). \quad (4.20)$$

In the general case $\vec{g} = g_y \vec{\tau}$ and $\dfrac{\partial \vec{g}}{\partial t} = \dfrac{\partial g_y}{\partial t}\vec{\tau} + g_y \dfrac{\partial \vec{\tau}}{\partial t}$. It's obvious that the supplement forces don't appear in the linear photon. In case the photon rolls up around any of the axis, which is perpendicular to the $y$-axis, we obtain the additional terms:

$$g_y \frac{\partial \vec{\tau}}{\partial t} = g_y \frac{v_s}{r_s}\vec{n} = g_y \omega_s \vec{n}, \quad (4.21)$$

Using Eqs. (4.19) and (4.20) for the photon $(E_z, H_z)$ we take the following normal force components:

$${}^z f_2 = \frac{1}{4\pi}\frac{\omega}{c} E_x H_z \vec{n} = \frac{1}{c} j_\tau \cdot H_z \vec{n} \quad (4.22)$$

$${}^z f_0 = \frac{1}{4\pi}\frac{\omega}{c} E_x^{\,2} = \frac{1}{c} j_\tau \cdot E_x, \quad (4.23)$$

for photon $(E_z, H_x)$ we take:

$${}^x f_2 = -\frac{1}{4\pi}\frac{\omega}{c} E_z H_x \vec{n} = -\frac{1}{c} j_\tau \cdot H_x \vec{n} \quad (4.24)$$

$${}^x f_0 = -\frac{1}{4\pi}\frac{\omega}{c} E_z^{\,2} = -\frac{1}{c} j_\tau \cdot E_z, \quad (4.25)$$

(here the upper left index shows the rotation axis $OZ$ or $OX$).
    As we see the Lorenz's force direction does not coincide with the $y$-axis, but is perpendicular to it.

**4.5. THE ELECTRODYNAMICS' FORM OF ELECTRON THEORY LANGRANGIAN**

    As it is known [2,9,10], the Lagrangian of Maxwell's theory on the free field is the first invariant of the theory $I_1$:

$$L_M \equiv I_1 = \frac{1}{8\pi}\left(\vec{E}^2 - \vec{H}^2\right), \quad (4.26)$$

As Lagrangian of Dirac's theory can be taken the expression [8]:

$$L_D = \psi^+\left(\hat{\varepsilon} + c\hat{\vec{\alpha}}\,\hat{\vec{p}} + \hat{\beta}\, m_e c^2\right)\psi, \quad (4.27)$$

For the wave along $y$-axis Eq. (4.27) can be written

$$L_D = \frac{1}{c}\psi^+ \frac{\partial \psi}{\partial t} - \psi^+ \hat{\alpha}_y \frac{\partial \psi}{\partial y} - i\frac{m_e c}{\hbar}\psi^+ \hat{\beta}\psi, \qquad (4.28)$$

Transfer the each term of (4.28) in electrodynamics' form we obtain for the twirled semi-photon Lagrangian:

$$L_s = \frac{\partial U}{\partial t} + div\,\vec{S} - i\frac{\omega_e}{8\pi}(\vec{E}^2 - \vec{H}^2), \qquad (4.29)$$

where $\omega_e = \frac{2mc^2}{\hbar}$; (note that we must differ the complex conjugate field vectors $\vec{E}^*$ and $\vec{E}$, $\vec{H}^*$ and $\vec{H}$).

The Eq. (4.29) can be written in other form. Using electrical and magnetic currents

$$j_\tau^e = i\frac{\omega_e}{4\pi}\vec{E} \qquad \text{and} \qquad j_\tau^m = i\frac{\omega_e}{4\pi}\vec{H}, \qquad (4.30)$$

we take:

$$L_s = \frac{\partial U}{\partial t} + div\,\vec{S} - \frac{1}{2}(\vec{j}_\tau^e \vec{E} - \vec{j}_\tau^m \vec{H}), \qquad (4.31)$$

Since $L_s = 0$ thanks to (4.1) we take the equation:

$$\frac{\partial U}{\partial t} + div\,\vec{S} - \frac{1}{2}(\vec{j}_\tau^e \vec{E} - \vec{j}_\tau^m \vec{H}) = 0, \qquad (4.32)$$

which is the energy-momentum conservation law equation with current.

According to (4.29) for the own electron electromagnetic wave the kind of Maxwell's equation Lagrangian exists, which differs from (4.26):

$$L_M = \frac{1}{8\pi}(\vec{E}^2 - \vec{H}^2) = \frac{i}{\omega_e}\left(\frac{\partial U}{\partial t} + div\,\vec{S}\right), \qquad (4.33)$$

### 4.6. ELECTROMAGNETIC FORM OF THE FREE ELECTRON EQUATION SOLUTION

In accordance with above results the electromagnetic form of the solution of the Dirac's free electron equation must be the twirled electromagnetic wave.

From the electron model point of view for the $y$-direction photon two solutions must exist: 1) for the wave, rotating around the $OZ$-axis

$$^{oz}\psi = \begin{pmatrix}\psi_1 \\ 0 \\ 0 \\ \psi_4\end{pmatrix} = \begin{pmatrix}E_x \\ 0 \\ 0 \\ iH_z\end{pmatrix}, \qquad (4.34)$$

and 2) for the wave, rotating around the $OX$-axis

$$^{ox}\psi = \begin{pmatrix}0 \\ \psi_2 \\ \psi_3 \\ 0\end{pmatrix} = \begin{pmatrix}0 \\ E_z \\ iH_x \\ 0\end{pmatrix}, \qquad (4.35)$$

Consider now the exact theory. It is known [8] that the solution of the Dirac's free electron equation (4.1) has the plane wave view:

$$\psi_j = b_j e^{i(\vec{k}\vec{r} - \omega t + \phi)}, \quad j = 1, 2, 3, 4, \qquad (4.36)$$

where the amplitudes $b_j$ are the numbers and $\phi$ is the initial wave phase. The functions (4.36) are the eigenfunctions of the energy-momentum operators, where $\hbar\omega$ and $\hbar\vec{k}$ are the energy-momentum eigenvalues. If we put the Eq. (4.36) in the Eq. (4.1) we obtain [9] the algebraic equation system, which are the homogenic linear equations for the $B_j = b_j e^{i\phi}$ values:

$$\begin{cases} (\varepsilon + m_e c^2)B_1 + cp_z B_3 + c(p_x - ip_y)B_4 = 0, \\ (\varepsilon + m_e c^2)B_2 + c(p_x + ip_y)B_3 - cp_z B_4 = 0, \\ (\varepsilon - m_e c^2)B_3 + cp_z B_1 + c(p_x - ip_y)B_2 = 0, \\ (\varepsilon - m_e c^2)B_4 + c(p_x + ip_y)B_1 - cp_z B_2 = 0, \end{cases} \quad (4.37)$$

where $\varepsilon = \hbar\omega$, $\vec{p} = \hbar\vec{k}$ are the numbers. This system has a solution only when the equation determinant is equal to zero: $(\varepsilon^2 - m_e^2 c^4 - c^2 \vec{p}^2)^2 = 0$. Here for each $\vec{p}$, the energy $\varepsilon$ has either positive value $\varepsilon_+ = +(c^2 \vec{p}^2 + m_e^2 c^4)^{\frac{1}{2}}$ or negative value $\varepsilon_- = -(c^2 \vec{p}^2 - m_e^2 c^4)^{\frac{1}{2}}$.

For $\varepsilon_+$ we have two linear-independent set of four orthogonal normalising spinors:

1) $B_1 = -\dfrac{cp_z}{\varepsilon_+ + m_e c^2}$, $B_2 = -\dfrac{c(p_x + ip_y)}{\varepsilon_+ + m_e c^2}$, $B_3 = 1$, $B_4 = 0$, \quad (4.38)

2) $B_1 = -\dfrac{c(p_x - ip_y)}{\varepsilon_+ + m_e c^2}$, $B_2 = \dfrac{cp_z}{\varepsilon_+ + m_e c^2}$, $B_3 = 0$, $B_4 = 1$, \quad (4.39)

and for $\varepsilon_-$:

3) $B_1 = 1$, $B_2 = 0$, $B_3 = \dfrac{cp_z}{-\varepsilon_- + m_e c^2}$, $B_4 = \dfrac{c(p_x + ip_y)}{-\varepsilon_- + m_e c^2}$, \quad (4.40)

4) $B_1 = 0$, $B_2 = 1$, $B_3 = \dfrac{c(p_x - ip_y)}{-\varepsilon_- + m_e c^2}$, $B_4 = -\dfrac{cp_z}{-\varepsilon_- + m_e c^2}$, \quad (4.41)

Let's analyse these solutions.

**At first**, the existing of two linear independent solutions corresponds with two independent orientation of the electromagnetic wave vectors (4.34) and (4.35) and gives the unique logic explanation for this fact.

**Secondly**, since $\psi = \psi(y)$, we have $p_x = p_z = 0$, $p_y = m_e c$ and for the field vectors we obtain: from Eqs. (4.38) and (5.39) for "positive" energy

$$B_+^{(1)} = \begin{pmatrix} 0 \\ b_2 \\ b_3 \\ 0 \end{pmatrix} \cdot e^{i\phi}, \quad B_+^{(2)} = \begin{pmatrix} b_1 \\ 0 \\ 0 \\ b_4 \end{pmatrix} \cdot e^{i\phi}, \quad (4.42)$$

and from (4.40) and (5.41) for "negative" energy

$$B_-^{(1)} = \begin{pmatrix} b_1 \\ 0 \\ 0 \\ b_4 \end{pmatrix} \cdot e^{i\phi}, \quad B_-^{(2)} = \begin{pmatrix} 0 \\ b_2 \\ b_3 \\ 0 \end{pmatrix} \cdot e^{i\phi}, \quad (4.43)$$

which correspond to (4.34) and (4.35).

**At third**, it is not difficult to calculate the correlation between components of the field vectors, putting $\varepsilon_+ = mc^2$ and $\phi = \dfrac{\pi}{2}$:

$$B_+^{(1)} = \begin{pmatrix} 0 \\ \dfrac{1}{2} \\ i \cdot 1 \\ 0 \end{pmatrix}, \quad B_+^{(2)} = \begin{pmatrix} -\dfrac{1}{2} \\ 0 \\ 0 \\ i \cdot 1 \end{pmatrix}, \quad (4.44)$$

And also for $\varepsilon_- = -mc^2$:

$$B_-^{(1)} = \begin{pmatrix} i \cdot 1 \\ 0 \\ 0 \\ -\dfrac{1}{2} \end{pmatrix}, \quad B_-^{(2)} = \begin{pmatrix} 0 \\ i \cdot 1 \\ \dfrac{1}{2} \\ 0 \end{pmatrix}, \quad (4.45)$$

Show the both orientations *a* and *b* of the initial photon on one picture (fig.3).

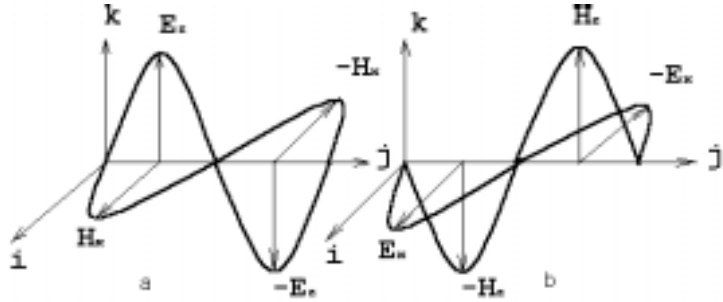

Fig.3. To Dirac's equation solution

As we can see, the first semi-periods on figs.3*a* and 3*b* correspond to the solutions $B_+^{(1)}$ and $B_+^{(2)}$. The second semi-periods correspond to the solutions $B_-^{(1)}$ and $B_-^{(2)}$. Note, that the appearance of minus sign before the electric field vector in $B_+^{(1)}$ and $B_-^{(1)}$ corrects automatically our mistake in (4.34).
   *The imaginary unit in the solutions indicates that the field vectors $\vec{E}$ and $\vec{H}$ are mutual orthogonal.*
   Father, from solution we can see that the magnetic field amplitude is two times less, than the electric field amplitude.

## 5. NON-LINEAR EQUATIONS AND THEIRS LAGRANGIANS

### 5.1. NON-LINEAR EQUATIONS OF TWIRLED SEMI-PHOTON

   The stability of twirled photon is possible only by the photons parts self-action. Obviously, the equation that describes the twirled semi-photon structure must be non-linear and like the equation, which describes photon-photon interaction [9,10].
   We could introduce the self-field interaction to the twirled semi-photon equation like the external field is introduced to the quantum [9,10] and classical [11] field equations (putting the photon mass equal to zero). But this equation may be obtained more rigorously.

For the electron the energy conservation equation is true:

$$\varepsilon^2 = c^2 \vec{p}^2 + m_e^2 c^2 , \qquad (5.1)$$

where $m_e$ is the electron mass and $\varepsilon$ and $\vec{p}$ are not operators in this case, but are the physical value of energy and momentum. The linear conservation law can be written then as:

$$\varepsilon = \pm \left( c\hat{\vec{\alpha}} \vec{p} + \hat{\beta} m_e c^2 \right), \qquad (5.2)$$

The relation (5.2) is rigorously, like as (5.1), since the second degree of (5.2) give the (5.1).

According to relationship (5.2) for the twirled semi-photon we can find:

$$\hat{\beta} m_e c^2 = - \left( \varepsilon_s - c\hat{\vec{\alpha}} \vec{p}_s \right),$$

Substitute Eq. (5.2) in Dirac's electron equation we obtain non-linear integral-differential equation:

$$\left[ \hat{\alpha}_0 (\hat{\varepsilon} - \varepsilon_s) + c\hat{\vec{\alpha}} ( \hat{\vec{p}} - \vec{p}_s) \right] \psi = 0 , \qquad (5.3)$$

which is, as we propose, the common form of the twirled semi-photon equation.

In the electromagnetic form we have:

$$\varepsilon_s = \int_{\Delta\tau} U \, d\tau = \frac{1}{8\pi} \int_{\Delta\tau} \left( \vec{E}^2 + \vec{H}^2 \right) d\tau, \qquad (5.4)$$

$$\vec{p}_s = \int_{\Delta\tau} \vec{g} \, d\tau = \frac{1}{c^2} \int_{\Delta\tau} \vec{S} \, d\tau = \frac{1}{4\pi} \int_{\Delta\tau} \left[ \vec{E} \times \vec{H} \right] d\tau, \qquad (5.5)$$

where $\Delta\tau$ is the semi-photon volume, which according to our model is equal approximate:

$$\Delta\tau \cong 2\pi^2 \zeta^2 r_s^3 , \qquad (5.6)$$

Using the quantum form of $U$ and $\vec{S}$:

$$U = \frac{1}{8\pi} \psi^+ \hat{\alpha}_0 \psi , \qquad (5.7)$$

$$\vec{S} = -\frac{c}{8\pi} \psi^+ \hat{\vec{\alpha}} \psi = c^2 \vec{g} , \qquad (5.8)$$

we can analyse the quantum form of (5.3). Since the Dirac's free electron equation solution is the plane wave, we have:

$$\psi = \psi_0 \, e^{i(\omega t - ky)} , \qquad (5.9)$$

Taking in to account (5.6) we can write (5.7) and (5.8) in next approximately form

$$\varepsilon_s = U \, \Delta\tau = \frac{\Delta\tau_s}{8\pi} \psi^+ \hat{\alpha}_0 \psi , \qquad (5.10)$$

$$\vec{p}_s = \vec{g} \, \Delta\tau = \frac{1}{c^2} \vec{S} \, \Delta\tau_s = -\frac{\Delta\tau_s}{8\pi c} \psi^+ \hat{\vec{\alpha}} \psi , \qquad (5.11)$$

Substitute (5.10) and (5.11) into (5.3) and taking in to account (5.6), we obtain the following approximately equation:

$$\frac{\partial \psi}{\partial t} - c\hat{\vec{\alpha}} \vec{\nabla}\psi + i\frac{\zeta^2}{2\alpha_q c} \cdot r_s^3 \left( \psi^+ \hat{\alpha}_0 \psi - \hat{\vec{\alpha}} \psi^+ \hat{\vec{\alpha}} \psi \right) \psi = 0 , \qquad (5.12)$$

It is not difficult to see that Eq. (5.12) is a non-linear equation of the same type, which was investigated by Heisenberg e.al. [9,12] and which played for a while the role of the unitary field theory equation. Moreover, Eq. (5.12) is obtained logically rigorously, contrary to the last one. Self-action constant $r_s$ [9] appears in Eq. (5.12) automatically and expresses the radius of the «bare» electron.

## 5.2. THE EQUATION OF TWIRLED SEMI-PHOTON MATTER MOTION

As it is known [8,10], the motion equations can be found from the next operator equation:

$$\frac{d\hat{L}}{dt} = \frac{\partial \hat{L}}{\partial t} + \frac{1}{i\hbar}\left(\hat{L}\hat{H} - \hat{H}\hat{L}\right), \qquad (5.13)$$

where $\hat{L}$ is the physical value operator, whose variation we want to find, and $\hat{H}$ is the Hamilton operator of Dirac's equation.

The Hamilton operator of the electron equation is equal:

$$\hat{H} = -c\hat{\vec{\alpha}}\,\hat{\vec{P}} - \hat{\beta}\,m_e c^2 + \varepsilon_e, \qquad (5.14)$$

where $\vec{P}$ is the full momentum of electron.

Let us $\hat{\vec{P}}_s$ be the full momentum of the twirled semi-photon. For $\hat{L} = \vec{P}_s$ from (5.14) substituting $\vec{v}_s = c\hat{\vec{\alpha}}$, where $\vec{v}_s$ - velocity of the semi-photon (electron) matter, we obtain:

$$\frac{d\vec{P}_s}{dt} = \left(\frac{\partial \vec{p}_s}{\partial t} + grad\ \varepsilon_s\right) - \left[\vec{v}_s \times rot\ \vec{p}_s\right], \qquad (5.15)$$

Since for the motionless electron $\dfrac{d\hat{\vec{P}}_s}{dt} = 0$, the motion equation is:

$$\left(\frac{\partial \vec{p}_s}{\partial t} + grad\ \varepsilon_s\right) - \left[\vec{v}_s \times rot\ \vec{p}_s\right] = 0, \qquad (5.16)$$

Passing to the energy and momentum densities

$$\vec{g}_s = \frac{1}{\Delta\tau_s}\vec{p}_s, \quad U_s = \frac{1}{\Delta\tau_s}\varepsilon_s, \qquad (5.17)$$

we obtain the equation of matter motion of twirled semi-photon:

$$\left(\frac{\partial \vec{g}_s}{\partial t} + grad\ U_s\right) - \left[\vec{v}_s \times rot\ \vec{g}_s\right] = 0 \qquad (5.18)$$

Let us analyse the physical means of Eq. (5.18), considering the motion equation of the ideal liquid in form of Lamb-Gromeka's equation [13]. In case the external forces are absent, this equation is:

$$\left(\frac{\partial \vec{g}_l}{\partial t} + grad\ U_l\right) - \left[\vec{v}_l \times rot\ \vec{g}_l\right] = 0, \qquad (5.19)$$

where $U_l, \vec{g}_l$ and $v_l$ are the energy, momentum density and velocity of ideal liquid.

Comparing (5.18) and (5.19) we see their mathematical identity. From this follows the interesting conclusion: the particle equation may be interpreted as the motion equation of ideal liquid.

According to (4.19) we have

$$\frac{\partial \vec{g}_s}{\partial t} + grad\ U_s = \vec{f}_L, \qquad (5.20)$$

where $f_L$ is the Lorenz force. As it is known the term $\left[\vec{v}_l \times rot\ \vec{g}_l\right]$ in Eg.(5.19) is responsible for centripetal acceleration. Probably, we have the same in (5.18). If the "photon liquid" moves along the ring of $r_s$ radius, then the angular motion velocity $\omega_s$ is tied with $rot\ \vec{v}_s$ by expression:

$$rot\ \vec{v}_s = 2\vec{\omega}_s = 2\omega_s \vec{e}_z^o,  \qquad (5.21)$$

and centripetal acceleration is:

$$\vec{a}_r = \frac{1}{2}\vec{v}_s \times rot\ \vec{v}_s = \frac{v^2}{r_s}\vec{e}_r^o = c\omega_s \vec{e}_r^o, \qquad (5.22)$$

where $\vec{e}_r^o$ is the unit radius-vector, $\vec{e}_z^o$ - is the unit vector of $OZ$-axis. As a result the equation (5.21) has the form of Newton's law:

$$\rho_m \vec{a}_r = \vec{f}_L, \qquad (5.23)$$

As a consequence of this analysis we can say that the twirled photons is like the vortex rings in ideal liquid.

### 5.3. LAGRANGIAN OF NON-LINEAR TWIRLED SEMI-PHOTON EQUATION

The Lagrangian of the non-linear equation is not difficult to obtain from Lagrangian of the linear Dirac's equation:

$$L_D = \psi^+\left(\hat{\varepsilon} + c\hat{\vec{\alpha}}\ \hat{\vec{p}} + \hat{\beta}\ m_e\ c^2\right)\psi, \qquad (5.24)$$

using the method by which we find the first degree non-linear equation.
Substitute (5.3) into (5.24) we obtain:

$$L_N = \psi^+\left(\hat{\varepsilon} - c\hat{\vec{\alpha}}\ \hat{\vec{p}}\right)\psi + \psi^+\left(\varepsilon_s - c\hat{\vec{\alpha}}\ \vec{p}_s\right)\psi, \qquad (5.25)$$

The expression (5.25) represents the common form of Lagrangian of the non-linear twirled semi-photon equation.
Using (5.10) and (5.11) we represent (5.25) in explicit quantum form:

$$L_N = i\hbar\left[\frac{\partial}{\partial t}\left[\frac{1}{2}(\psi^+\psi)\right] - c\ div(\psi^+\hat{\vec{\alpha}}\psi)\right] + \frac{\Delta\tau_s}{8\pi}\left[(\psi^+\psi)^2 - (\psi^+\hat{\vec{\alpha}}\psi)^2\right], \qquad (5.26)$$

Normalising $\psi$-function we obtain:

$$L_N = \frac{1}{8\pi\ m_e c}L_N, \qquad (5.27)$$

Using Eqs. (5.4) and (5.5) and transforming Eq.(5.25) in electrodynamics' form we find:

$$L_N = i\frac{\hbar}{2m_e c^2}\left(\frac{\partial U}{\partial t} + div\ \vec{S}\right) + \frac{\Delta\tau}{m_e c^2}\left(U^2 - c^2\vec{g}^2\right), \qquad (5.28)$$

Here accordingly (4.29) the first summand may be replaced through

$$i\frac{\hbar}{2mc^2}\left(\frac{\partial U}{\partial t} + div\ \vec{S}\right) = \frac{1}{8\pi}\left(\vec{E}^2 - \vec{H}^2\right), \qquad (5.29)$$

Now we'll transform the second summand, using the follow known *transformation (which is the electrodynamics' form of Fierz's correlation):*

$$(8\pi)^2\left(U^2 - c^2\vec{g}^2\right) = \left(\vec{E}^2 + \vec{H}^2\right)^2 - 4\left(\vec{E} \times \vec{H}\right)^2 = \left(\vec{E}^2 - \vec{H}^2\right)^2 + 4\left(\vec{E} \cdot \vec{H}\right)^2, \qquad (5.30)$$

So we have:

$$L_N = \frac{1}{8\pi}\left(\vec{E}^2 - \vec{H}^2\right) + \frac{\Delta\tau}{(8\pi)^2 m_e c^2}\left[\left(\vec{E}^2 - \vec{H}^2\right)^2 + 4\left(\vec{E}\cdot\vec{H}\right)^2\right], \quad (5.31)$$

As we see, the Lagrangian of the non-linear twirled semi-photon equation contains only the invariant of Maxwell's theory. Accordingly to our conception the Eq. (5.31) is Lagrangian of the semi-photon itself interaction. In quantum mechanics [9,10] is known the Lagrangian of photon-photon interaction:

$$L_{p-p} = \frac{1}{8\pi}\left(\vec{E}^2 - \vec{H}^2\right) + b\left[\left(\vec{E}^2 - \vec{H}^2\right)^2 + 7\left(\vec{E}\cdot\vec{H}\right)^2\right] + ..., \quad (5.32)$$

where $b = \frac{2}{45}\frac{e^4 \hbar}{m_e^4 c^7}$. Comparing (5.31) to (5.32) it is not difficult to see, that they coincide up to the digital coefficients. This confirm our hypothesis that electron is an interacting photon itself.

Let's now analyse the quantum form Lagrangian. The Eq. (5.25) can be written in form:

$$L_N = \psi^+ \hat{\alpha}_\mu \partial_\mu \psi + \frac{\Delta\tau}{8\pi}\left[\left(\psi^+ \hat{\alpha}_0 \psi\right)^2 - \left(\psi^+ \hat{\vec{\alpha}}\, \psi\right)^2\right], \quad (5.33)$$

The Eq. (5.30) has the following quantum form:

$$\left(\psi^+ \hat{\alpha}_0 \psi\right)^2 - \left(\psi^+ \hat{\vec{\alpha}}\, \psi\right)^2 = \left(\psi^+ \hat{\alpha}_4 \psi\right)^2 + \left(\psi^+ \hat{\alpha}_5 \psi\right)^2, \quad (5.34)$$

Then from (5.33) we obtain:

$$L_N = \psi^+ \hat{\alpha}_\mu \partial_\mu \psi + \frac{\Delta\tau_s}{8\pi}\left[\left(\psi^+ \hat{\alpha}_4 \psi\right)^2 - \left(\psi^+ \hat{\alpha}_5 \psi\right)^2\right], \quad (5.35)$$

It is not difficult to see that Lagrangian (5.35) practically (if we use the $\gamma$-set of Dirac's matrices instead of $\alpha$-set) coincide with the Nambu's and Jona-Losinio's Lagrangian [14], which is the Lagrangian of the relativistic superconductivity theory and gives the solution to the problem of the appearance of the elementary particles mass by the mechanism of the vacuum symmetry spontaneous breakdown.

Note that in our theory, the breakdown of symmetry also takes place when a mass of particles appears (it corresponds to the Cooper's pair decay on electron and «hole» in the superconductivity theory).

### 6.0  PHSYCAL SENSE OF THE PARTICULARITIES OF DIRAC'S EQUATION

The Dirac equation has a lot of particularities. In modern interpretation these particularities are considered as entirely mathematical features that don't have a physical meaning. For example, according to [15] (section 34-4) "one can prove that all the physical consequences of Dirac's equation do not depend on the special choice of Dirac's matrices. They would be the same if a different set of four 4x4 matrices with the specification, $\hat{\alpha}_\mu \cdot \hat{\alpha}_\nu = \begin{cases} 1, & \text{if } \mu = \nu \\ 0, & \text{if } \mu \neq \nu \end{cases}$ had been chosen. In particular it is possible to interchange the roles of the four matrices by unitary transformation. So, their differences are only apparent".

The mathematical properties of Dirac's matrices are well known: they are anticommutative and hermitian; they build a group of 16 matrices; the bilinear forms of these matrices have definite transformation properties, which correspond to the vector properties of the (e.g.) electrodynamics; repeatedly it was pointed that in the classical physics these matrices describe the rotations, etc.[16].

We have shown above that *the Dirac matrices describe the electromagnetic vectors moving along the curvilinear trajectories (not vectors in the curvilinear space!).* From this suggestion follows the interpretation of all matrix properties.

Below we will consider how the Dirac matrices are joined with the electromagnetic Dirac's equation forms. We will show that all the electron Dirac's equation particularities have the exact electrodynamics meaning. We will also answer to the question, why the "choice of Dirac's matrices" exists, but the results "don't depend on the special choice". Here we will also analyse the physical sense of the different form of Dirac's equation in regard to various Dirac's equation matrix representation.

### 6.1   PYSICAL SENSE OF THE DIFFERENT FORMS OF DIRAC'S EQUATION

As it is known, there are two Dirac's equation forms:

$$[(\hat{\alpha}_o \hat{\varepsilon} + c\hat{\vec{\alpha}}\ \hat{\vec{p}}) + \hat{\beta}\ mc^2]\psi = 0, \qquad (6.1)$$

$$\psi^+[(\hat{\alpha}_o \hat{\varepsilon} - c\hat{\vec{\alpha}}\ \hat{\vec{p}}) - \hat{\beta}\ mc^2] = 0, \qquad (6.2)$$

which correspond to the two signs of the relativistic energy expression:

$$\varepsilon = \pm\sqrt{c^2 \vec{p}^2 + m^2 c^4}, \qquad (6.3)$$

Also, as it is known that, for each sign of equation (6.3) there are two hermitian-conjugate Dirac's equations.

Here we consider the physical meaning of all these equations.

Let us consider at first two hermitian-conjugate equations, corresponding to the minus sign in (6.3):

$$[(\hat{\alpha}_o \hat{\varepsilon} + c\hat{\vec{\alpha}}\ \hat{\vec{p}}) + \hat{\beta}\ mc^2]\psi = 0, \qquad (6.4)$$

$$\psi^+[(\hat{\alpha}_o \hat{\varepsilon} + c\hat{\vec{\alpha}}\ \hat{\vec{p}}) + \hat{\beta}\ mc^2] = 0, \qquad (6.5)$$

Using (4.2) from (6.4) and (6.5) we obtain:

$$\begin{cases} \dfrac{1}{c}\dfrac{\partial E_x}{\partial t} - \dfrac{\partial H_z}{\partial y} + i\dfrac{\omega}{c}E_x = 0, \\[4pt] \dfrac{1}{c}\dfrac{\partial E_z}{\partial t} + \dfrac{\partial H_x}{\partial y} + i\dfrac{\omega}{c}E_z = 0, \\[4pt] \dfrac{1}{c}\dfrac{\partial H_x}{\partial t} + \dfrac{\partial E_z}{\partial y} - i\dfrac{\omega}{c}H_x = 0, \\[4pt] \dfrac{1}{c}\dfrac{\partial H_z}{\partial t} - \dfrac{\partial E_x}{\partial y} - i\dfrac{\omega}{c}H_z = 0, \end{cases} \qquad (6.8)$$

$$\begin{cases} \dfrac{1}{c}\dfrac{\partial E_x}{\partial t} - \dfrac{\partial H_z}{\partial y} - i\dfrac{\omega}{c} E_x = 0, \\[4pt] \dfrac{1}{c}\dfrac{\partial E_z}{\partial t} + \dfrac{\partial H_x}{\partial y} - i\dfrac{\omega}{c} E_z = 0, \\[4pt] \dfrac{1}{c}\dfrac{\partial H_x}{\partial t} + \dfrac{\partial E_z}{\partial y} + i\dfrac{\omega}{c} H_x = 0, \\[4pt] \dfrac{1}{c}\dfrac{\partial H_z}{\partial t} - \dfrac{\partial E_x}{\partial y} + i\dfrac{\omega}{c} H_z = 0, \end{cases} \qquad (6.9)$$

As we see, the equation (6.8) and (6.9) are the Maxwell equations with complex currents that we name the twirled semi-photon equations. As we see the equations (6.8) and (6.9) are differed by the current directions. We could foresee this result before the calculations, since the functions $\psi^+$ and $\psi$ are differed by the argument signs:

$$\psi^+ = \psi_0 e^{-i\omega t} \quad \text{and} \quad \psi = \psi_0 e^{i\omega t}.$$

Let us compare now the equations corresponding to the both plus and minus signs of (6.3). For the plus sign of (2.2) we have following two equations:

$$\left[ \left( \hat{\alpha}_o \hat{\varepsilon} - c\hat{\vec{\alpha}}\ \hat{\vec{p}} \right) - \hat{\beta}\, mc^2 \right] \psi = 0, \qquad (6.10)$$

$$\psi^+ \left[ \left( \hat{\alpha}_o \hat{\varepsilon} - c\hat{\vec{\alpha}}\ \hat{\vec{p}} \right) - \hat{\beta}\, mc^2 \right] = 0, \qquad (6.11)$$

The electromagnetic form of the equation (2.10) is:

$$\begin{cases} \dfrac{1}{c}\dfrac{\partial E_x}{\partial t} + \dfrac{\partial H_z}{\partial y} + i\dfrac{\omega}{c} E_x = 0, \\[4pt] \dfrac{1}{c}\dfrac{\partial E_z}{\partial t} - \dfrac{\partial H_x}{\partial y} + i\dfrac{\omega}{c} E_z = 0, \\[4pt] \dfrac{1}{c}\dfrac{\partial H_x}{\partial t} - \dfrac{\partial E_z}{\partial y} - i\dfrac{\omega}{c} H_x = 0, \\[4pt] \dfrac{1}{c}\dfrac{\partial H_z}{\partial t} + \dfrac{\partial E_x}{\partial y} - i\dfrac{\omega}{c} H_z = 0, \end{cases} \qquad (6.12)$$

Obviously, the electromagnetic form of equation (6.11) will have the opposite signs of the currents compared to (6.12).

Comparing (6.12) and (6.8) we see that the equation (6.12) can be considered as Maxwell's equation in the left co-ordinate system. So as not to use the left co-ordinate system, together with the wave function of the electron $\psi_{ele}$ we can consider the wave function of positron in the form, which correspond to the right co-ordinate system:

$$\psi_{pos} = \begin{pmatrix} E_x \\ -E_z \\ iH_x \\ -iH_z \end{pmatrix}, \qquad (6.13)$$

Then, contrary to the system (6.12) we get the system (6.9). The transformation of the function $\psi_{ele}$ to the function $\psi_{pos}$ is named the charge conjugation operation.

Note that the electron and positron wave functions can be considered as the retarded and advanced waves. So the above result links also with the theory of advanced waves of Wheeler and Feynman [17]. (See also Dirac's work on time-symmetric classical electrodynamics [18] and about this theme Konopinski's book [19].

## 6.2 PHYSICAL SENSE OF THE MATRIX CHOICE

Now we will consider the bispinor form of Dirac's equation from the point of view that it describes the twirled object.

Consider the Dirac bispinor equation (6.1)-(6.2) with $\alpha$ - set of Dirac's matrices, as above.

As we saw above this matrix sequence $(\hat{\alpha}_1, \hat{\alpha}_2, \hat{\alpha}_3)$ agrees to the twirled semi-photon having $y$-direction. But herewith only the $\hat{\alpha}_2$-matrix is "working", and other two matrices don't give the terms of the equation. The verification of this fact is the Poynting vector calculation: the bilinear forms of $\hat{\alpha}_1, \hat{\alpha}_3$-matrices are equal to zero, and only the matrix $\hat{\alpha}_2$ gives the right non-zero component of Poynting's vector.

The question arises how to describe the photons having initially $x$ and $z$ - directions? It is not difficult to see [15] that the matrices' sequence is not determined by the some special requirements. In fact, this matrices' sequence can be changed without breaking some quantum electrodynamics results.

So we can write three groups of matrices, each of which corresponds to the one and only one direction, introducing the axes' indexes, which indicate the photon direction:

$$(\hat{\alpha}_{1x}, \hat{\alpha}_{2y}, \hat{\alpha}_{3z}), \quad (\hat{\alpha}_{2x}, \hat{\alpha}_{3y}, \hat{\alpha}_{1z},), \quad (\hat{\alpha}_{2z}, \hat{\alpha}_{1y}, \hat{\alpha}_{3x}).$$

Let us choose now the wave function forms, which give the correct Maxwell equations. We will take as initial the form for the $y$ - direction, which we already used, and from them, by means of the indexes' transposition around the circle (see fig.4) we will get other forms for $x$ and $y$ - directions.

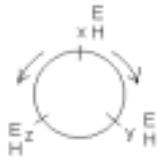

Fig.4

Since in this case the Poynting vector has the minus sign, we can suppose that the transposition takes place counterclockwise.

Let us check the Poynting vector values:

1) $\psi = \psi(y)$, $(\hat{\alpha}_{1x}, \hat{\alpha}_{2y}, \alpha_{3z})$, $\psi = \begin{pmatrix} E_x \\ E_z \\ iH_x \\ iH_z \end{pmatrix}$, $\psi^+ = \begin{pmatrix} E_x & E_z & -iH_x & -iH_z \end{pmatrix}$. (6.14)

$\psi^+ \hat{\alpha}_{1x} \psi = 0$, $\psi^+ \hat{\alpha}_{2y} \psi = -2[\vec{E} \times \vec{H}]_y$, $\psi^+ \hat{\alpha}_{3z} \psi = 0$.

2) $\psi = \psi(x)$, $(\hat{\alpha}_{2x}, \hat{\alpha}_{3y}, \hat{\alpha}_{1z})$, $\psi = \begin{pmatrix} E_z \\ E_y \\ iH_z \\ iH_y \end{pmatrix}$, $\psi^+ = \begin{pmatrix} E_z & E_y & -iH_z & -iH_y \end{pmatrix}$, (6.15)

$\psi^+ \hat{\alpha}_{2x} \psi = -2[\vec{E} \times \vec{H}]_x$, $\psi^+ \hat{\alpha}_{3y} \psi = 0$, $\psi^+ \hat{\alpha}_{1z} \psi = 0$.

3) $\psi = \psi(z)$, $(\hat{\alpha}_{2z}, \hat{\alpha}_{1y}, \hat{\alpha}_{3x})$, $\psi = \begin{pmatrix} E_y \\ E_x \\ iH_y \\ iH_x \end{pmatrix}$, $\psi^+ = \begin{pmatrix} E_y & E_x & -iH_y & -iH_x \end{pmatrix}$, (6.16)

$\psi^+ \hat{\alpha}_{3x} \psi = 0$, $\psi^+ \hat{\alpha}_{1y} \psi = 0$, $\psi^+ \hat{\alpha}_{2z} \psi = -2[\vec{E} \times \vec{H}]_z$

As we see, we took the correct result: by the counterclockwise indexes' transposition the wave functions describe the photons, which are moved in negative directions of the corresponding co-ordinate axes.

We may hope that by the clockwise indexes' transposition, the wave functions will describe the photons, which are moved in positive directions of co-ordinate axes. Prove this:

1) $\psi = \psi(y)$, $(\hat{\alpha}_{1x}, \hat{\alpha}_{2y}, \alpha_{3z})$, $\psi = \begin{pmatrix} E_z \\ E_x \\ iH_z \\ iH_x \end{pmatrix}$, $\psi^+ = \begin{pmatrix} E_z & E_x & -iH_z & -iH_x \end{pmatrix}$, (6.17)

$\psi^+ \hat{\alpha}_{1x} \psi = \begin{pmatrix} E_z & E_x & -iH_z & -iH_x \end{pmatrix} \begin{pmatrix} iH_x \\ iH_z \\ E_x \\ E_z \end{pmatrix} = iE_z H_x + iE_x H_z - iE_x H_z - iE_z H_x = 0$,

$\psi^+ \hat{\alpha}_{2y} \psi = \begin{pmatrix} E_z & E_x & -iH_z & -iH_x \end{pmatrix} \begin{pmatrix} H_x \\ -H_z \\ -iE_x \\ iE_z \end{pmatrix} = E_z H_x - E_x H_z - E_x H_z + E_z H_x =$

$= 2(E_z H_x - E_x H_z) = 2[\vec{E} \times \vec{H}]_y$

$\psi^+ \hat{\alpha}_{3z} \psi = \begin{pmatrix} E_z & E_x & -iH_z & -iH_x \end{pmatrix} \begin{pmatrix} iH_z \\ -iH_x \\ E_z \\ -E_x \end{pmatrix} = iE_z H_z - iE_x H_x - iE_z H_z + iE_x H_x = 0$.

2) $\psi = \psi(x)$, $(\hat{\alpha}_{2x}, \hat{\alpha}_{3y}, \hat{\alpha}_{1z})$, $\psi = \begin{pmatrix} E_y \\ E_z \\ iH_y \\ iH_z \end{pmatrix}$, $\psi^+ = \begin{pmatrix} E_y & E_z & -iH_y & -iH_z \end{pmatrix}$, (6.18)

$$\psi^+\hat{\alpha}_{2_x}\psi = 2[\vec{E}\times\vec{H}]_x, \quad \psi^+\hat{\alpha}_{3_y}\psi = 0, \quad \psi^+\hat{\alpha}_{1_z}\psi = 0.$$

3) $\psi = \psi(z)$, $(\hat{\alpha}_{2_z}, \hat{\alpha}_{1_y}, \hat{\alpha}_{3_x})$, $\psi = \begin{pmatrix} E_x \\ E_y \\ iH_x \\ iH_y \end{pmatrix}$, $\psi^+ = (E_x \ E_y \ -iH_x \ -iH_y)$, (6.19)

$$\psi^+\hat{\alpha}_{3_x}\psi = 0, \quad \psi^+\hat{\alpha}_{1_y}\psi = 0, \quad \psi^+\hat{\alpha}_{2_z}\psi = 2[\vec{E}\times\vec{H}]_z$$

As we see, we also get the correct result.
   Now we will prove that the above choice of the matrices give the correct electromagnetic equation forms. Using for example the bispinor Dirac's equation (6.10)

$$[(\hat{\alpha}_o\hat{\varepsilon} - c\hat{\vec{\alpha}}\,\hat{\vec{p}}) - \hat{\beta}\,mc^2\,]\psi = 0$$

and transposing the indexes clockwise we obtain for the positive direction photons the following results:
1) for $x$-direction:

$$\frac{1}{c}\frac{\partial}{\partial t}\begin{pmatrix} E_y \\ E_z \\ iH_y \\ iH_z \end{pmatrix} + \frac{\partial}{\partial x}\begin{pmatrix} H_z \\ -H_y \\ -iE_z \\ iE_y \end{pmatrix} = -i\frac{mc}{\hbar}\begin{pmatrix} E_y \\ E_z \\ -iH_y \\ -iH_z \end{pmatrix}, \quad \text{or} \quad \begin{cases} \frac{1}{c}\frac{\partial E_y}{\partial t} + \left(\frac{\partial H_z}{\partial x}\right) = -i\frac{\omega}{c}E_y, & a^* \\ \frac{1}{c}\frac{\partial E_z}{\partial t} - \left(\frac{\partial H_y}{\partial x}\right) = -i\frac{\omega}{c}E_z, & a' \\ \frac{1}{c}\frac{\partial H_y}{\partial t} - \left(\frac{\partial E_z}{\partial x}\right) = i\frac{\omega}{c}H_y, & a'' \\ \frac{1}{c}\frac{\partial H_z}{\partial t} + \left(\frac{\partial E_y}{\partial x}\right) = i\frac{\omega}{c}H_z, & a^{**} \end{cases}$$

(6.20)
2) for $y$-direction:

$$\frac{1}{c}\frac{\partial}{\partial t}\begin{pmatrix} E_z \\ E_x \\ iH_z \\ iH_x \end{pmatrix} + \frac{\partial}{\partial y}\begin{pmatrix} H_x \\ -H_z \\ -iE_x \\ iE_z \end{pmatrix} = -i\frac{mc}{\hbar}\begin{pmatrix} E_z \\ E_x \\ -iH_z \\ -iH_x \end{pmatrix}, \quad \text{or} \quad \begin{cases} \frac{1}{c}\frac{\partial E_z}{\partial t} + \left(\frac{\partial H_x}{\partial y}\right) = -i\frac{\omega}{c}E_z, & b^* \\ \frac{1}{c}\frac{\partial E_x}{\partial t} - \left(\frac{\partial H_z}{\partial y}\right) = -i\frac{\omega}{c}E_x, & b' \\ \frac{1}{c}\frac{\partial H_z}{\partial t} - \left(\frac{\partial E_x}{\partial y}\right) = i\frac{\omega}{c}H_z, & b'' \\ \frac{1}{c}\frac{\partial H_x}{\partial t} + \left(\frac{\partial E_z}{\partial y}\right) = i\frac{\omega}{c}H_x, & b^{**} \end{cases}$$

(6.21)

2) for $z$-direction:

$$\frac{1}{c}\frac{\partial}{\partial t}\begin{pmatrix} E_x \\ E_y \\ iH_x \\ iH_y \end{pmatrix} + \frac{\partial}{\partial z}\begin{pmatrix} H_y \\ -H_x \\ -iE_y \\ iE_x \end{pmatrix} = -i\frac{mc}{\hbar}\begin{pmatrix} E_x \\ E_y \\ -iH_x \\ -iH_y \end{pmatrix}, \quad \text{or} \quad \begin{cases} \frac{1}{c}\frac{\partial E_x}{\partial t} + \left(\frac{\partial H_y}{\partial z}\right) = -i\frac{\omega}{c}E_x, & c^* \\ \frac{1}{c}\frac{\partial E_y}{\partial t} - \left(\frac{\partial H_x}{\partial z}\right) = -i\frac{\omega}{c}E_y, & c' \\ \frac{1}{c}\frac{\partial H_x}{\partial t} - \left(\frac{\partial E_y}{\partial z}\right) = i\frac{\omega}{c}H_x, & c'' \\ \frac{1}{c}\frac{\partial H_y}{\partial t} + \left(\frac{\partial E_x}{\partial z}\right) = i\frac{\omega}{c}H_y, & c^{**} \end{cases} \quad (6.22)$$

So we have obtained three equation groups, each of which contains four equations (totally, there are 12 equations), as it is necessary for the description of all wave directions.

In the same way for all the other Dirac's equation forms the analogue results can be obtained.

Obviously, it is possible via canonical transformations to choose the Dirac matrices so that the initial photon could have not only $x$, $y$ or $z$ directions, but any other direction.

### 6.2 THE ELECTRODYNAMICS SENSE OF CANONICAL TRANSFORMATIONS

The choice of the matrixes $\hat{\alpha}$, made by us, is not unique. As it is known there is the canonical transformation of a kind
$$\hat{\alpha} = S\hat{\alpha}'S, \quad (6.23)$$
where $S$ is a unitary matrix which consists of four lines and four rows, and $\hat{\alpha}'$ are the new matrices. The following substitution for new functions corresponds to the above transformation:
$$\psi = S\psi', \quad (6.24)$$
For example, if we choose $\alpha'$ matrices as:

$$\hat{\alpha}'_1 = \begin{pmatrix} 0 & 1 & 0 & 0 \\ 1 & 0 & 0 & 0 \\ 0 & 0 & 0 & 1 \\ 0 & 0 & 1 & 0 \end{pmatrix}, \quad \hat{\alpha}'_2 = \begin{pmatrix} 0 & -i & 0 & 0 \\ i & 0 & 0 & 0 \\ 0 & 0 & 0 & i \\ 0 & 0 & i & 0 \end{pmatrix}, \quad \hat{\alpha}'_3 = \begin{pmatrix} 1 & 0 & 0 & 0 \\ 0 & -1 & 0 & 0 \\ 0 & 0 & 1 & 0 \\ 0 & 0 & 0 & -1 \end{pmatrix},$$

$$\vec{\alpha}'_4 = \begin{pmatrix} 0 & 0 & 0 & -1 \\ 0 & 0 & 1 & 0 \\ 0 & 1 & 0 & 0 \\ -1 & 0 & 0 & 0 \end{pmatrix}, \quad \hat{\alpha}'_5 = \begin{pmatrix} 0 & 0 & 0 & -i \\ 0 & 0 & i & 0 \\ 0 & -i & 0 & 0 \\ i & 0 & 0 & 0 \end{pmatrix}; \quad (6.25)$$

the functions $\psi$ are connected to the functions $\psi'$ by the relationships:
$$\psi = \frac{\psi'_1 - \psi'_4}{\sqrt{2}}, \quad \psi = \frac{\psi'_2 + \psi'_3}{\sqrt{2}}, \quad \psi = \frac{\psi'_1 + \psi'_4}{\sqrt{2}}, \quad \psi = \frac{\psi'_2 - \psi'_3}{\sqrt{2}}, \quad (6.26)$$
The unitary matrix, which corresponds to this transformation, is:

$$S = \frac{1}{\sqrt{2}} \begin{Bmatrix} 1 & 0 & 0 & -1 \\ 0 & 1 & 1 & 0 \\ 1 & 0 & 0 & 1 \\ 0 & 1 & -1 & 0 \end{Bmatrix}, \qquad (6.27)$$

It is not difficult to check, that all these relations have a certain sense in the electromagnetic theory. Using (6.6) and (6.26) we have for $\psi'$- functions the following relationships:

$$\frac{\psi'_1 - \psi'_4}{\sqrt{2}} = E_x, \quad \frac{\psi'_2 + \psi'_3}{\sqrt{2}} = E_z, \quad \frac{\psi'_1 + \psi'_4}{\sqrt{2}} = iH_x, \quad \frac{\psi'_2 - \psi'_3}{\sqrt{2}} = iH_z, \qquad (6.28)$$

whence:

$$\psi' = \begin{pmatrix} \psi'_1 \\ \psi'_2 \\ \psi'_3 \\ \psi'_4 \end{pmatrix} = \frac{\sqrt{2}}{2} \begin{pmatrix} (E_x + iH_x) \\ (E_z + iH_z) \\ (E_z - iH_z) \\ (E_x - iH_x) \end{pmatrix}, \qquad (6.28)$$

By usual substitution it is possible to check up, that this function gives the correct bilinear forms and Maxwell's equations.

Thus, the matrix choice changes only the mathematical description, not the physical results of the theory. On the other hand, the knowledge of the electrodynamics properties of the Dirac matrices can be very useful. So if we knew this fact, e.g., in early times of development of the week interaction theory (i.e. the Fermi theory), the 16 Dirac matrix control wouldn't be needed, since only the vector and pseudovector matrices $\hat{\vec{\alpha}}$ and $\hat{\alpha}_5$ give the electrodynamics invariant (see above).

**CONCLUSION**

Let us enumerate some results of our theory, which explain the features of the modern quantum theory:

**A principle of uncertainty**: as it is known, the uncertainty relation arises in the theory of any wave. In our theory the particles are represented by the twirled electromagnetic waves. Therefore this principle is fair for elementary particles.

**An operational method**: the operators in quantum physics arise as operators of the equations of (twirled) electromagnetic waves.

**Wave packages**: at early stages of development of the quantum mechanics an almost classical explanation was found for many features of the quantum mechanics on the base of the wave packages. The only disadvantage of wave packages - their dispersion - is absent in the twirled waves thanks to their monochromaticity and a harmonicity.

**Statistical interpretation of wave function**: if an electron represents an EM wave, the square of $\psi$-function will describe the density of energy of the EM wave, referred to rest energy of the electron.

**The phase and gauge invariancy** plays a basic role in the modern theory of elementary particles. As it is known [20] "gauge invariancy means physically the same, as the phase invariancy". In case that elementary particles are the electromagnetic waves, the phase invariancy represents the construction tool of the theory of elementary particles.

**The dualism a wave - particle** ceases be a riddle in the theory of the twirled photons: elementary particles really represent simultaneously both waves and particles.

**Etc.**

**Characteristics of elementary particles.**

Above we had the opportunity to be convinced that in our theory these characteristics appear consistently and logically:

**The rest mass of particles.** The masses of the particles arise as the "stopped" electromagnetic energy of a photon. Only the photon rest mass is equal to zero; all other particles should have a rest mass.

**The electric charge** appears as the consequence of the occurrence of a tangential displacement current of Maxwell at a twirling of a photon. The value of a charge according to our theory is defined only by some constants.

**Universality of the charge**: according to our theory the charge is defined only by means of twirling of the semi-photons, which is the same for all Universe.

**The ambiguity of a charge** corresponds to the interaction of circular currents (by observance of Pauli's principle)

**The law of the electric charge conservation** follows from the way of formation of charges by the division of the whole photon into two twirled half periods: electron and a positron. The quantities of "positive" and "negative" half-periods are equal in the Universe.

**Infinite problems**: in the above theory the charge and mass infinite problems don't exist.

**The spin of particles** arises owing to the twirling of the field of an electromagnetic wave.

**Helicity** is defined by poloidal rotation of a field of particles.

**The division of particles on bosons and fermions**: bosons are the twirled photons and fermions are the twirled semi-photons. As other opportunities of division of a wave do not exist, there is not any other type of particles in the nature.

**The existence of particles and antiparticles** corresponds to the antisymmetry of the twirled semi-photons.

**The spontaneous breakdown of symmetry of vacuum and the occurrence of the particle mass** is connected with the change of symmetry of a linear photon at the moment of its twisting and division into two half-periods.

**The Zitertbewegung** corresponds to the rotation of the semi-photon fields.

**Etc.**

Many facts of the formal structure of Dirac's equations do not find an explanation in the traditional quantum mechanics. In the above research simple and convincing explanations to all features of Dirac's equations were received, for example:
1. Why the bispinor equation contains four equations?
2. Why the Dirac's matrices in the classical theory describe the vector rotations in the space?
3. Why these matrices have the certain electrodynamics vector properties and why do 16 matrices exist?
4. Why the Dirac's equations are received from Klein-Gordon's equation?
5. Why the theory of groups (i.e. the theory of the space rotation and other transformations of symmetry) is a basis for the search of the physical theory invariants?
6. The sense of the equivalence of Dirac's equation forms, as the equations of the waves of various space directions.
7. Etc.

**REFERENCES**

1. R.F. O'Connel. Does the electron have a structure? Foundation of Physics, **23**, 487-429 (1993).
2. M.-A. Tonnelat. Les Principes de la Theorie Electromagnetique et de la Relativite. Masson et C. Editeurs, Paris, 1959.
3. E.Schroedinger. Verwachene Eigenwertspektra. Berl.Ber. (Phys.-math. Kl.), 1929, S. 668-682.
4. V.Fock. Natshala kvantovoi mechaniki (in Russ.). KUBUTSH, L., 1932.
5. W.Heisenberg. Physical principles of quantum theory. 1930
6. H.Stanley Allen. The angular momentum of ring electron. Phil.Mag., Ser.6, v.41, No. 241, p.113 (1921)
7. I.M.Matora. Relativistic theory of charge circulation in electron. Hadronic Journal v. 20, pp. 147, 267, 523 (1997)
8. L.T. Schiff. Quantum Mechanics (2nd edition, Mc-Graw-Hill Book Company, Jnc, New York, 1955).
9. A.Sokoloff, D.Iwanenko. Kwantowaja teorija polja (in Russian). Moskwa-Leningrad, 1952.
10. A.I. Achieser, W.B. Berestetski. Quantum electrodynamics. Moscow, 1969.
11. J.W. Leech: Classical Mechanics. Methuen and Co. Ltd., London, Willy and Sons Jnc., New York 1958.
12. W. Heisenberg. Introduction to the Unified Field Theory of Elementary Particles. Interscience Publishers, London, New York, Sydney, 1966.
13. Lamb H, Lehrbuch der Hydrodynamik. 2 Aufl., Leipzig, B.G.Treubner, 1931
14. J. Nambu and G. Jona-Losinio. Phys. Rev., **122**, No.1, 345-358 1961); **124**, 246 (1961).
15. Fermi, E. *Notes on quantum mechanics*. The University of Chicago press, 1960.
16. Ryder, L.H. *Quantum field theory*. Cambridge University press, Cambridge, 1985.
17. Wheeler, J.A., and Feynman, R.P. Reviews of Modern Physics, 17, 157 (1945); Wheeler, J.A. Reviews of Modern Physics 29,463 (1957).
18. Dirac, P.A. M. Proc. Royal Soc. London A167, 148 (1938)
19. Konopinski, E.J. *Electromagnetic Fields and Relativistic Particles*, Chapter XIV, (McGraw-Hill, New York, 1980.
20. Yang, Ch. Uspekhi Fizicheskikh Nauk, v.132, N.1 (1980).